\documentclass[12pt,preprint]{aastex}

\slugcomment{To appear in Astrophysical Journal June 10, 2008, v680n 1 issue }
\shorttitle{Dynamics of the Solar Magnetic Network. II}
\shortauthors{Hasan \& van Ballegooijen}

\begin{document}

\title{Dynamics of the Solar Magnetic Network.\\
II. Heating the magnetized chromosphere}

\author{S. S. Hasan}
\affil{Indian Institute of Astrophysics, Bangalore-560034, India}
\email{hasan@iiap.res.in}
\and
\author{A. A. van Ballegooijen}
\affil{Harvard-Smithsonian Center for Astrophysics,
60 Garden Street, Cambridge, MA~02138, U.S.A.}

\begin{abstract}
We consider recent observations of the chromospheric network, and
argue that the bright network grains observed in the Ca~II H \& K
lines are heated by an as yet unidentified quasi-steady process. We
propose that the heating is caused by dissipation of short-period
magnetoacoustic waves in magnetic flux tubes (periods less than
100~s). Magnetohydrodynamic (MHD) models of such waves are presented.
We consider wave generation in the network due to two separate
processes: (a) by transverse motions at the base of the flux tube;
and (b) by the absorption of acoustic waves generated in the ambient
medium. We find that the former mechanism leads to an efficient
heating of the chromosphere by slow magnetoacoustic waves propagating
along magnetic field lines. This heating is produced by shock waves
with a horizontal size of a few hundred kilometers. In contrast,
acoustic waves excited in the ambient medium are converted into
transverse fast modes that travel rapidly through the flux tube and do
not form shocks, unless the acoustic sources are located within 100 km
from the tube axis. We conclude that the magnetic network may be
heated by magnetoacoustic waves that are generated in or near the flux
tubes.
\end{abstract}


\section{Introduction}

In the chromosphere on the quiet Sun it is useful to distinguish
between the magnetic network on the boundary of supergranulation
cells \citep[][]{Simon1964}, where strong magnetic fields are
organized in mainly vertical flux tubes, and internetwork regions
in the cell interiors, where magnetic fields are weaker and
dynamically less important.

The canonical picture of the magnetic network is that it consists of
vertical magnetic fields clumped into elements or flux tubes that are
located in intergranular lanes, have magnetic field strengths in the
kilogauss range, and have diameters of the order of 100 km or less at
their footpoints in the photosphere \citep[e.g.,][]{Gaizauskas1985,
Zwaan1987}. These magnetic elements can be identified with bright
points seen in images taken in the G-band (4305 {\AA}).
High resolution observations show that these flux elements are
in a highly dynamical state due to buffeting by convective flows on
granular and supergranular scales \citep[e.g.,][]{Muller1994,
Berger1995, Berger1998, Berger1996, Nisenson2003}. With the
availability of new ground-based telescopes at excellent sites and
sophisticated image reconstruction techniques it is now possible to
examine magnetic elements with an improved resolution of about
0.17 arcsec and investigate their structure and dynamics in
unprecedented detail \cite[e.g.,][]{Berger2004, Rouppe2005,
Langangen2007}. High-quality observations of photospheric magnetic
fields are now also being obtained with the Solar Optical Telescope
(SOT) on Hinode \citep[e.g.,][]{Lites2007, Lites2008, Rezaei2007}.
Magneto-convection models have been developed to understand the
three-dimensional (3-D) structure and evolution of the magnetic field
and its interactions with convective flows \citep[e.g.,][]{Vogler2005,
Schaffenberger2006}.

The chromospheric network is most clearly seen in filter images taken
in the Ca II H \& K lines \citep[e.g.,][]{Gaizauskas1985, Rutten2007}
and in the Ca II IR triplet \citep[][]{Cauzzi2007}. In H or K line
images the network shows up as a collection of ``coarse mottles'' or
``network grains'' that stand out against the darker background.
The network grains are continuously bright with intensities that vary
slowly in time, in contrast to the ``fine mottles'' or ``cell grains''
which are located in the cell interiors and are much more dynamic
\citep[e.g.,][]{Rutten1991}. In space-time diagrams derived from
Ca II H spectra \citep[e.g., Fig.~3 in][]{Cram1983}, the network
grains show up as long bright streaks, indicating lifetimes of at
least 10 min. The Ca~II H line profiles of network grains are more
symmetrical than those from the cell interior, and both the
${\rm H}_{2V}$ and ${\rm H}_{2R}$ features generally appear bright 
\citep[][]{Lites1993, Sheminova2005}. Superposed on this constant
brightness are low-frequency oscillations with periods of 5 to 20
min. \citep[][]{Lites1993, Kontogiannis2006, Tritschler2007};
these oscillations have also been observed at UV wavelengths
\citep[e.g.,][]{Bloomfield2006}.

The physical processes that produce the enhanced emission in Ca~II
network grains have not been identified. Are they heated by wave
dissipation, and if so, what is the nature of these waves? In this
paper we focus on the question why the emission from network grains is
so constant in time. We discuss the observations, and investigate
whether waves with periods $P < 100$ s can provide the heating of the
grains.

\citet[][hereafter Paper I]{Hasan2005} considered (magneto)acoustic
waves generated at the interface of magnetic flux tubes and the
outside medium. They suggested that such waves are generated as a
result of interactions of flux tubes with turbulent downflows in
intergranular lanes just below the photosphere. Intergranular lanes
have widths of order 100 km, and downflow velocities of several km s$^{-1}$,
so velocity fluctuations $\delta v \sim 1$ km s$^{-1}$ may be expected.  Such
fluctuations would produce transverse displacements of the flux tubes
within the intergranular lanes. Assuming random displacements with
amplitude $\delta x \sim 10$ km, the typical time scale is $\tau \sim
10$ s. Such perturbations would produce transverse waves with period
$P \sim 2 \tau$ that travel upward along the flux tube, as well as
pressure fluctuations on either side of the tube (Paper~I). Therefore,
from a theoretical perspective it is reasonable to expect the
existence of short-period magnetoacoustic waves in flux tubes
($P \sim 20$ s).

In this paper we present models of wave propagation in magnetic flux
tubes that are relevant to the heating of the network grains.
Specifically, we present two-dimensional (2-D) magnetohydrodynamic
(MHD) simulations of the interactions of short-period waves with
magnetic flux sheets (we shall use the terminology flux sheet and flux
tube interchangeably).  Although such simulations do not include the
Alfv\'{e}n mode, they provide a possible model for the waves involved
in heating the Ca II grains. This work is a continuation of the study
in Paper~I.

The plan of the paper is as follow. In \S 2 we review recent
observations of the chromospheric network, and we argue that the
network grains are heated by an as yet unidentified quasi-steady
process. We propose that the heating is due to short-period waves
($P < 100$ s), although there is little evidence for such waves at
present. In \S 3 we describe a 2-D MHD model for the propagation
of short-period waves in magnetic flux tubes. In \S 4 we present
simulation results for waves generated {\it inside} a flux tube;
interaction of waves on different flux tubes are also considered.
In \S 5 we consider the interaction of a flux tube with waves
generated by acoustic sources located {\it outside} the flux tube.
The final \S 6 discusses and summarizes the main results of this
investigation.

\section{Observations of Ca II Network Grains}

\begin{figure}
\epsscale{.80}
\plotone{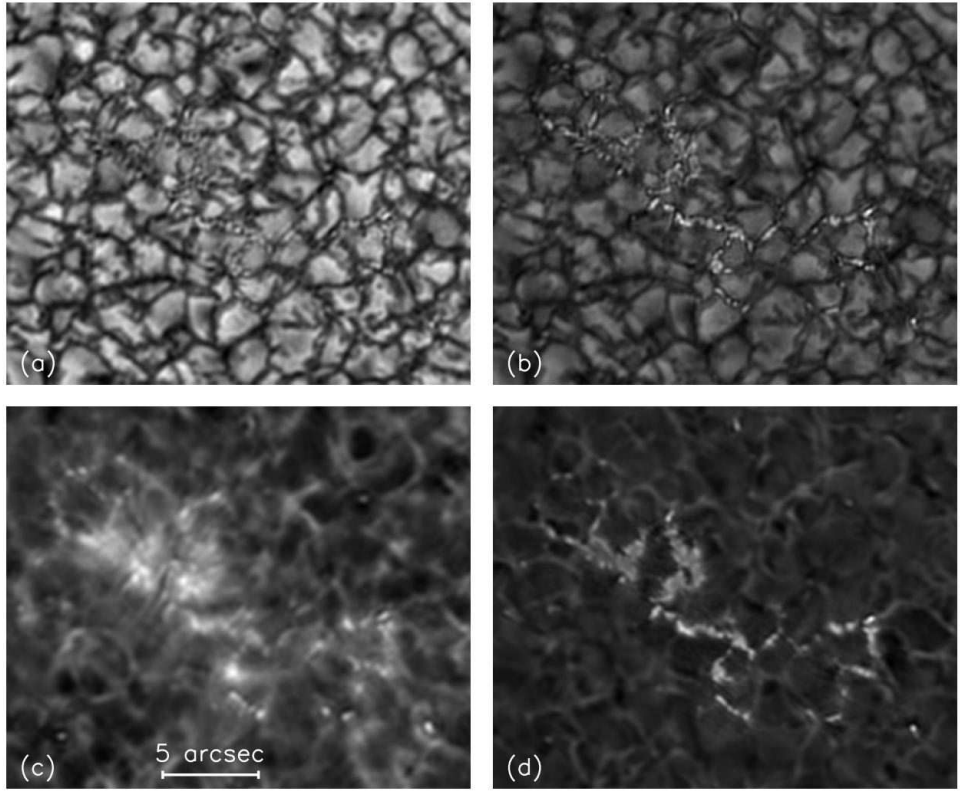}
\caption
{Dutch Open Telescope (DOT) observations of a magnetic network patch
in the solar photosphere and chromosphere. The images were taken at
solar disk center on 2007 April 12 at 9:35 UT. (a) blue continuum;
(b) G band at 4305 {\AA}; (c) Ca II H line center; (d) Ca II H offband
(-2.35 {\AA}). The network patch consists of many small flux elements
of sub-arcsecond size.
The bright network grains seen in the Ca~II H line center image
coincide with G-band bright points, indicating that the chromospheric
heating is localized directly above the photospheric flux elements.
The Ca~II network grains are more diffuse than the G-band bright
points, which may be due to the spreading of flux tubes with height.}
\label{fig1}
\end{figure}

Figure \ref{fig1} shows an example of a network element observed with
the Dutch Open Telescope \citep[DOT,][]{Rutten2005}. The different
panels show filtergrams in the blue continuum, G band, Ca II H line
center, and Ca II H off band.  The DOT uses a filter with a 1.4 {\AA}
FWHM passband centered on Ca II H (3968 {\AA}), which covers the two
chromospheric emission peaks on either side of line center, but also
includes parts of the inner wings of the line, which originate in the
upper photosphere.  An examination of the G-band image confirms the
picture of a network patch consisting of multiple flux elements. A
comparison with the Ca II line center image shows that the excess
chromospheric emission is localized directly above the photospheric
flux tubes, although the bright features seen in Ca II are more
diffuse than those seen in the G-band. Time sequences of such images
show that the chromospheric network is continually bright
\citep[e.g.,][]{Tritschler2007}.

\citet{Sheminova2005} combined high-resolution Ca II H and K spectra
of network elements with flux tube modeling, and found that the excess
brightness in the wings of the H and K lines compared with the quiet
photosphere is primarily due to low density, not to mechanical
heating. However, both the ${\rm H}_{2V}$ and ${\rm H}_{2R}$ emission
features appear bright, indicating sustained non-thermal heating at
heights where these features are formed \citep[also see][]{Lites1993}.
The Ca~II profiles from network elements are nearly symmetric,
indicating that the heating does not involve strong shocks such as
those found in internetwork regions \citep[][]{Carlsson1997}.

\citet{Rutten2006, Rutten2007} presented reviews and a synthesis of
recent high-resolution observations of the solar chromosphere. In the
Ca II H \& K lines, the network shows up as a collection of ``Ca II
bright points'' or ``grains''. He also identifies an exciting new
phenomenon called ``straws'' that extend radially outward from network
bright points \citep[see Figure 8 in][]{Rutten2007}. These straws are
very thin, occur in ``hedge rows'', and are very short-lived (10-20 s).
They appear to be closely related to the so-called type-II spicules
recently identified in limb observations with Hinode/SOT \citep[][]
{DePontieu2007b}. The type-II spicules are much more dynamic than
ordinary (type-I) spicules, are very thin (width $\sim 100$ km), have
lifetimes of 10-150 s, and seem to be rapidly heated to transition
region temperatures, sending material though the chromosphere at
speeds of 50-150 $\rm km ~ s^{-1}$. De Pontieu and collaborators
suggest that type-II spicules may be due to small-scale reconnection
events in the chromosphere. \citet{DePontieu2007c} show that the
spicules exhibit transverse motions with velocities of 10 to 25
$\rm km ~ s^{-1}$, which provides evidence for the existence of
Alfv\'{e}n waves in the chromosphere.

In Ca II observations of network elements on the solar disk, the thin
straws are superposed on the bright grains. According to
\citet{Rutten2006}: ``The network bright points are less sharp than and
differ in morphology from G-band bright points ... bright stalks
emanate a few arcseconds from them at this resolution ... diffuse Ca
II H core brightness spreads as far or farther" (see his Fig.~2).
Rutten conjectures that the bright Ca II network grains are nothing
but a collection of straws seen nearly end-on. However, this is
difficult to reconcile with the observed steadiness of the emission
from network elements at high spatial resolution. Since the straws are
very short-lived and dynamic, a superposition of many such features
would be required to produce apparantly steady emission, but such high
density of straws is not consistent with the limb observations by
\citet{DePontieu2007b, DePontieu2007c}. Also, if the plasma in the
straws is hot ($T \sim 10^5$ K), as \citet{Rutten2006} suggests, one
would expect the straws to produce pure emission profiles, but the
observed Ca II H \& K line profiles from network elements are nearly
symmetric and always show a central reversal \citep[see Fig.~5 in][]
{Sheminova2005}.  Therefore, we suggest that most of the emission from
Ca II grains is due to some quasi-steady heating process at heights of
500 km to 1500 km inside the magnetic flux tubes, and is not
associated with spicule-like activity at larger heights.

Our interpretation of the Ca II observations is summarized in Figure
\ref{fig2}, where we show a vertical cross section of a
magnetic network element consisting of several discrete flux tubes.
We suggest that the Ca II network grains are located inside the
magnetic flux tubes, and give rise to the bulk of the Ca II emission
from the network element. The grains are thought to be located at
heights between 500 km and 1500 km above the photosphere where the
flux tubes are no longer ``thin'' compared to the pressure scale
height (about 200 km), but are still well separated from each other.
The Ca II straws (type-II spicules) have widths of order 100 km,
and are located at larger heights (several Mm) where the widths of the
flux tubes are much larger than 100 km. Therefore, we suggest the
straws are not directly associated with the network grains in the low
chromosphere. In Figure \ref{fig2} we assumed that the straws
(type-II spicules) are located at the {\it interfaces} between the
flux tubes, as suggested by \citet{vanB+Nisenson1998}.

Observations have shown ``magnetic shadows'' surrounding network
elements where the oscillations in the 2 - 4 minute range are
suppressed \citep[][]{Judge2001, Krijger2001, Vecchio2007}. These
shadows have been attributed to the effects of highly structured
magnetic fields on wave propagation and mode conversion in the
chromosphere \citep[][]{McIntosh2001}. Network elements on the quiet
Sun exhibit oscillations with periods in the range 5 - 20 min.
\citep[e.g.,][]{Lites1993, Bloomfield2006}. These long-period
oscillations are thought to be driven by the interaction of vertical
or inclined magnetic flux tubes with global p-modes and convective
flows in the photosphere, producing magnetoacoustic shocks that drive
spicules and spicule-like features at larger heights in the
chromosphere \citep[][]{DePontieu2004, DePontieu2005, Hansteen2006,
Heggland2007, Rouppe2007}. \citet{Jefferies2006} argued that inclined
magnetic fields at the boundaries of supergranules may act as portals
through which low-frequency magnetoacoustic waves can propagate into
the solar chromosphere \citep[also see][] {McIntosh2006}.

It seems unlikely that long-period waves are also responsible for the
{\it heating} of the Ca~II network grains in the low chromosphere.
Simulations of shock waves with periods $P \sim 200$ s in a
plane-parallel, non-magnetic atmosphere have shown that such waves
produce large asymmetries in the Ca~II H line profiles, and strong
variations in the integrated emission. If the grains were heated by
such long-period waves, they should exhibit similar strong intensity
variations. This is not observed, so the long-period waves observed
in network elements cannot be the main source of heating for the Ca~II
grains.

Network grains could possibly be heated by dissipation of waves with
shorter periods ($P < 100$ s). Ground-based observations of
high-frequency waves in small network elements are affected by seeing,
so it is possible that waves with periods $P < 100$ s do exist in
network elements but are simply not observable from the ground.
This hypothesis could be tested using Ca~II H images from Hinode/SOT
\citep[][] {Tsuneta2007, Kosugi2007}, keeping in mind of course that
the passband of the Ca~II H filter also includes a significant
photospheric contribution.  Alternatively, network grains could be
heated by dissipation of Alfv\'{e}n waves. There is evidence for
Alfv\'{e}n waves in spicules \citep[][] {DePontieu2007c}, and if these
waves originate below the photosphere they must travel through the
heights where the Ca~II network grains are located. At present there
is no direct observational evidence for Alfv\'{e}n waves in flux
elements in the photosphere, nor at heights where the Ca~II H line is
formed. Also, simulations of such waves in flux tubes would require
3-D MHD models. Therefore, in the present paper we focus on
magnetoacoustic waves, which can be simulated with 2-D MHD models.

\section{Models for Magnetoacoustic Waves in Network Elements}

\subsection{Background}

Earlier work on wave dynamics idealized the network in terms of thin
flux tubes \citep[e.g.,][]{Roberts1978}, and treated wave propagation
in terms of the well known transverse (kink) and longitudinal
(sausage) modes \citep[e.g.,][]{Spruit1981}. Several investigations
have focused on the generation and propagation of transverse and
longitudinal wave modes and their dissipation in the chromosphere
\citep[e.g.,][and references therein]{Zhugzhda1995, Fawzy2002,
Ulmschneider2003}.  \citet{Hasan1999} examined the excitation of
transverse and longitudinal waves in magnetic flux tubes by the impact
of fast granules on flux tubes, as observed by \citet{Muller1992} and
\citet{Muller1994}, and following the investigation by
\citet{Choudhuri1993}, who studied the generation of kink waves by
footpoint motion of flux tubes. The observational signature of the
modelled process was highly intermittent in radiation emerging in the
H and K lines, contrary to observations. By adding waves that were
generated by high-frequency motion due to the turbulence of the medium
surrounding flux tubes the energy injection into the gas inside a flux
tube became less intermittent, and the time variation of the emergent
radiation was in better agreement with the more steady observed
intensity from the magnetic network \citep[][]{Hasan2000}.

The above studies on the magnetic network make use of two important
idealizations: they assume that the magnetic flux tubes are thin,
an approximation that becomes questionable at about the height of
formation of the emission peaks in the cores of the H and K lines;
and they neglect the interaction of neighboring flux tubes. Some
progress on the first issue was made by \citet{Rosenthal2002} and
\citet{Bogdan2003}, who studied wave propagation in a 2-D
stratified atmosphere, assuming the magnetic field can be approximated
by a potential field.  They examined the propagation of waves that
are excited from spatially localized sources in the
photosphere, and found that there is strong {\it mode coupling}
between fast and slow waves at the so-called magnetic canopy, which
they identify as the surface where the magnetic and gas pressures are
comparable. However, potential-field models do not provide an accurate
description of the boundary layers between the flux tubes and the
ambient medium, where electric currents flow perpendicular to the
magnetic field. In Paper~I we considered more realistic, magnetostatic
equilibrium models in which flux sheets are embedded in a field-free
medium. In such models the boundary layer between the flux tube and
its surroundings is sharper than in a potential field model (the width
of the boundary layer is spatially resolved, see \S 3.2). This
strongly affects the propagation and reflection of waves at the
boundary \citep[][]{Hasan2006}. Also, the shape of the $\beta=1$
surface in a magnetostatic model can be quite different from that
in a potential-field model.

In the present paper we further develop the model for MHD wave
propagation presented in Paper~I. The initial configuration consists
of one or more flux sheets in magnetostatic equilibrium with a
relatively sharp interface between the flux sheets and the surrounding
gas, unlike previous studies in which a potential field is considered
\citep[e.g.,][]{Rosenthal2002, Bogdan2003}. The present work goes
beyond Paper~I in three important aspects: firstly, it seeks to
understand how the dynamics of a flux tube is affected when the
magnetic field in the wave excitation region is not that strong,
so that the $\beta = 1$ surface (approximately the level where the
sound and Alfv\'en speeds become equal) lies above the base on the
tube axis.  Observations have revealed that the magnetic field
strengths in the network are not uniformly kilogauss but have a range
of values that go from strong to moderate \citep[e.g.,][]{Berger2004}.
For moderate field strengths, wave generation occurs in a region where
the field is essentially weak. Such waves undergo transformation
higher up in the atmosphere.  Secondly, we allow for the presence of
multiple flux tubes in a network patch. We investigate the dynamical
coupling between neighboring flux tubes and identify the regions
responsible for the enhanced emission in the magnetic chromosphere.
Finally, we examine the hypothesis that the magnetic network can be
heated by acoustic waves from the ambient medium. More details of the
model are provided in the next subsections.

\citet{Hansteen2006} and \citet{DePontieu2007a} performed advanced
2-D numerical simulations that span the entire solar atmosphere from
the convection zone to the lower corona. These simulations describe
the evolution of a radiative MHD plasma: the model includes the
effects of non-gray, non-LTE radiative energy transport in the
photosphere and chromosphere, as well as field-aligned thermal
conduction and optically thin radiative losses in the transition region
and corona. They find that MHD waves generated by convective flows and
oscillations in the lower photosphere naturally leak upward into the
magnetized chromosphere, where they form shocks that can drive
spicule-like jets. These models include more physics than the one
presented here, but we believe our approach is adequate for the task
of simulating short-period magnetoacoustic waves.

\subsection{Equilibrium Model}

In the solar photosphere the magnetic flux tubes are spatially
separated from each other, and are rooted in intergranular lanes. In
Paper I we considered a model for the structure of a magnetic network
element as a collection of unipolar flux tubes \citep[also see][]
{Cranmer2005, Cranmer2007}. We presented a two-dimensional model
with an array of identical flux tubes separated by 1200 km. The flux
tubes expand with height and merge into a monolithic structure at a
height of about 500 km. In the present paper we present modified
versions of this model. The method for constructing such models is
summarized below.

The equilibrium magnetic field ${\bf B} (x,z)$ is a function of
horizontal coordinate $x$ and height $z$ above the level where the
optical depth $\tau_{5000} = 1$ in the photosphere. The field is
assumed to be periodic in the $x$ coordinate; the period length
$L_x$ equals the distance between the flux tubes. The tubes are
assumed to symmetric with respect to their vertical axes, therefore
in the equilibrium model only one half of one flux tube needs to be
considered. For the model presented in \S 4.1, the size of the
computational domain is given by $\Delta x = L_x / 2 = 1800$ km and
$\Delta z = 2950$ km, and we use a grid spacing of 5 km. At the
lower boundary ($z = 0$), the flux tube has a nearly Gaussian profile
with $1/e$-width of 65 km. The 2-D structure of the flux tube is
determined by solving the force balance equation:
\begin{equation}
- \nabla p + \rho {\bf g} + \frac{1}{4\pi} ( \nabla \times {\bf B} )
\times {\bf B} = 0 , \label{eq:force}
\end{equation}
where ${\bf g} = -g \hat{\bf z}$ is the gravitational acceleration,
$p(x,z)$ is the gas pressure, and $\rho(x,z)$ is the density. The
third term describes the Lorentz force due to electric currents at
the boundary between the flux tube and its surroundings. The magnetic
field is written in terms of a flux function $A(x,z)$ such that $B_x
= - \partial A / \partial z$ and $B_z = \partial A / \partial x$.
The gas pressure is given by
\begin{equation}
p(x,z) = p_{int}(z) \left[ 1 + \beta_0^{-1} F(A) \right] ,
\end{equation}
where $p_{int} (z)$ is the internal gas pressure as function of height
along the axis of the flux tube ($A = 0$); $\beta_0$ is the ratio of
gas- and magnetic pressures at the base (on axis); and $F(A)$ is a
function describing the variation of gas pressure across field lines
($F=0$ on axis and $F=1$ in the ambient medium). A similar expression
holds for the gas density. The functions $p_{int} (z)$ and $F(A)$ are
described in Paper I, except that in the present paper we use $\beta_0
= 2$. Equation (\ref{eq:force}) is solved for $A(x,z)$ using an energy
minimization technique (see Paper I). Table 1 lists the basic
parameters of the equilibrium model on the tube axis and in the
ambient medium at the base ($z = 0$) and at height $z = 1500$ km in
the chromosphere.

To construct the initial conditions for the MHD dynamical model, the
equilibrium model is mirrored horizontally to create a single flux
tube, and then further replicated to produce an array of flux tubes.
Only the lower part of the equilibrium model is used ($z < 1600$ km).
Therefore, the magnetic field at the top of the MHD model is not
vertical.

\subsection{Dynamical Model and Boundary Conditions}

We consider wave generation in the configuration described in the
previous section by perturbing the lower boundary and solving the
2-D MHD equations in conservation form for an inviscid adiabatic
fluid.  These consist of the usual continuity, momentum, entropy
(without sources) and magnetic induction equations \citep[see][]
{Steiner1994}. The unknown variables are the density, momentum,
entropy per unit mass and the magnetic field.  We assume that the
plasma consists of fully ionized hydrogen with a mean molecular weight
of 1.297. The temperature is computed from the specific entropy and
the pressure is found using the ideal gas law.

The MHD equations are solved following the numerical procedure given
by \citet{Steiner1994} and also described in \citet{Hasan2005}.
Briefly, the equations are discretized on a 2-D mesh and solved using
a finite difference method based on the flux-corrected transport (FCT)
scheme of \citet{Oran1987}. The time integration is explicit and has
second-order accuracy in the time step. Small time steps are required
to satisfy the Courant condition in the upper part of the domain,
where the Alfv\'en speed is large.

Periodic boundary conditions are used at the horizontal boundaries.
At the top boundary, (a) the vertical and horizontal components of
momentum are set to zero; (b) the density is determined using linear
logarithmic extrapolation; (c) the horizontal component of the
magnetic field and temperature are set equal to the corresponding
values at the preceding interior point. The vertical component of the
magnetic field is determined using the condition $\nabla \cdot
{\bf B} = 0$. Similar conditions are used at the lower boundary,
except for the density, temperature and horizontal component of the
velocity (or momentum).  The density is obtained using cubic spline
extrapolation, the temperature is kept constant at its initial value
and the horizontal or vertical velocity is specified at the lower
boundary.

\subsection{Limitations of the Model}

Before describing the simulation results, it is appropriate to mention
some of the limitations of the present investigation. Our model does
not include the convection zone, therefore the wave excitation
mechanism is somewhat idealized. Also, we have used an adiabatic
energy equation, thereby neglecting radiative losses. The top boundary
conditions cause wave reflection, so we can run the simulations only
for a short period of time (about 180 s). Also, with the present code
we are limited in our choice for the height of the upper boundary;
for larger heights the Alfv\'{e}n speed near the top becomes too large
and the time step too small to make useful simulations.

Our analysis is based on a 2-D treatment in which the Alfv\'{e}n wave
is absent, and the coupling of the Alfv\'{e}n mode to the slow and
fast modes is neglected. Furthermore, in three dimensions the acoustic
waves generated by transverse motions of a flux tube can travel
azimuthally around the flux tube, which is not possible in a 2-D flux
sheet. Therefore, the pressure fluctuations $\delta p$ surrounding a
3-D tube should be smaller than those near a 2-D flux sheet. However,
based on a simple model for waves generated by an oscillating rigid
cylinder in a uniform medium, we estimate that the pressure
fluctuations are not significantly reduced compared to the 3-D case
when the wave period $P < 2 \pi R /C_s$, where $R$ is the flux tube
radius and $C_s$ is the sound speed in the ambient medium. For
$R = 100$ km and $C_s = 7.1$ $\rm km ~ s^{-1}$ we require $P < 88$ s,
so our approximation of the flux tubes as 2-D sheets is adequate for
short-period waves, but would not be accurate for longer wave
periods. In this paper we consider waves with a period of 24 s.

\section{Results for Wave Generation Inside the Flux Sheets}

\subsection{Dynamics of a single flux sheet} 

We first consider wave generation in a single flux sheet driven by
transverse periodic motions at the lower boundary.  Compared to
Paper~I we consider a lower value of the magnetic field strength of
1000 G on the axis at the base $z=0$. In this case, the $\beta=1$
surface occurs above the base within the sheet as opposed to Paper~I
where the field at the lower boundary is 1500 G (on the axis). The
reason for treating this case is the evidence from observations
\citep[][]{Berger2004} for the existence of network elements with
moderate field strengths. We examine the wave pattern that develops
in a flux tube due to a transverse motion at the lower boundary given
by the following form:
\begin{equation} v_x(x,0,t)=v_0\sin (2\pi t/P), \label{a1}
\end{equation} 
where $v_0$ denotes the amplitude of the horizontal motion and $P$ the
wave period. Similar to Paper I, we choose $v_0 = 750$ m s$^{-1}$ and
$P=24$~s.  Figures \ref{fig3}a and \ref{fig3}c shows the temperature
perturbation $\Delta T$ (the temperature difference  with respect to
the initial value at the same location) at 75~s, 122~s and 153~s. The
black curves denote the magnetic field lines and the white curves
depict constant $\beta$ contours with values 0.1 (upper curve), 1.0
(thick curve) and 10.0 (lower curve). 
 
The horizontal motions at the lower boundary produce compressions and
decompressions of the gas in the flux tube which generate an acoustic
like wave (most effectively at the interface between the tube and
ambient medium as shown in Paper I) that propagates isotropically with
an almost constant sound speed.  This can be discerned by the almost
constant spacing in the semicircular color pattern. In the central
section of the tube, a transverse slow MHD wave is generated that is
essentially guided along the field lines.  At $t=75$~s, we find from
Figure \ref{fig3}a that the wave pattern is confined below the
$\beta=1$ surface ($z=0.5$~Mm). In this region, where the magnetic
field can be regarded as weak, the acoustic (fast) mode travels ahead
of the (slow) MHD wave (which travels at the Alfv\'en speed).  On the
tube axis the acoustic and Alfv\'en speeds at the base are 7.1 km
s$^{-1}$ and 5.6 km s$^{-1}$ whereas at $z=500$~km they are 7.3 km
s$^{-1}$ and 8.0 km s$^{-1}$. At the $\beta = 1$ level there is a
strong coupling between the two modes \citep[as previously
demonstrated by][]{Rosenthal2002, Bogdan2003, Hasan2005}.

Figures \ref{fig3}b and \ref{fig3}c depict the situation at instants
where the wave has crossed the $\beta = 1$ surface. Above this level,
the acoustic wave (which is now a slow mode propagating along the
field) continues to travel at about 8 km s$^{-1}$ but the MHD wave
(fast mode) is speeded up due to the rapid increase of the Alfv\'en
speed with height. The faint blue and yellow halos in the flux tube
above the last semi-circle shows the fast mode. It should also be
noted that as the acoustic-like waves move upwards, their amplitude
increases due to the decrease in density with height resulting in
compressional heating. From Figures \ref{fig3}b and \ref{fig3}c we
find temperature enhancements of around 900 K in the upper atmosphere
due to amplitude increase of the acoustic wave amplitude.

A comparison of the present calculations for a network flux element
with a field on the axis of about 1000~G (corresponding to $\beta =
2$) with the case treated in Paper I of a stronger field of 1500 G
($\beta = 0.5$) shows that there are several qualitative similarities
in both cases.  These include the acoustic pattern in the ambient
medium and weak field regions ($\beta \gg 1$) as well as the strong
heating in the central regions of the tubes.  The main difference is
in the character of the waves near the tube axis: in Paper I, the
field in the central regions is strong at all heights, whereas in the
present case, the field becomes strong only above $z=0.5$~Mm leading
to a layer where strong mode coupling occurs. It is worthwhile to
clarify what happens to a wave when it crosses this layer. As pointed
out by \citet{Cally2007}, a fast wave in a high $\beta$ region
(essentially acoustic in character) propagating along a field line
changes labels from fast to slow as it crosses the $\beta = 1$ (or
more accurately the one on which the sound and Alfv\'en speeds are
equal, but this difference is not significant for the purpose of the
present analysis). In the $\beta < 1$ region, this wave is now a slow
mode, but still acoustic in nature. This phenomenon is essentially
{\em mode transmission}. However, the incident fast wave can also be
transformed to a magnetic mode (a fast wave in the magnetized
region). This corresponds to a {\em mode conversion}. In practice,
when the wave propagation vector makes a finite angle with respect to
the magnetic field lines, both mode {\em transmission} and {\em
conversion} occur.

Thus, in Paper I the wave generated in the tube due to footpoint
motion is a fast transverse MHD mode that continues to be fast (near
the axis) as it propagates upwards. On the other hand, in the present
case, the footpoint motions produce a slow transverse MHD wave that is
transmitted as a fast MHD (still transverse) as it crosses the
$\beta=1$ layer. There is, however, no mode conversion.

\subsection{Dynamics of multiple flux sheets} 

Let us now extend the analysis to a situation where we have multiple
flux sheets. This allows us to incorporate the influence of
neighboring flux tubes on each other. Figure \ref{fig4} shows the wave
pattern arising due to a transverse motion of the lower boundary where
the horizontal component of the velocity has the form given by
equation (\ref{a1}) with the same parameters as before. The flux tubes
in Figure \ref{fig4} are identical to each other and have a field
distribution on the base that is the same as in Figure \ref{fig3}.
However, above the base the field lines do not flare out as much as in
the previous case, but straighten out at a lower height due to the
effect of adjacent flux tubes. At $t=75$~s, the waves in the
individual tubes are sufficiently well separated from each other and
the wave pattern in each tube is qualitatively similar to the case
treated in \S 4.1. However, at $t = 122$~s waves emanating from
neighboring tubes interact with each other especially in the ambient
medium.  However, the wave pattern in any tube is not significantly
affected by the presence of its neighbors. Furthermore, the slow
magnetoacoustic waves above the $\beta = 1$ surface are confined close
to the central regions of the tubes, where they steepen and produce
enhanced heating. This heating appears to be dominantly caused by the
wave motions generated at the footpoints and {\em not} by the
penetration of acoustic waves from the ambient medium or by waves
coming from neighboring tubes (we shall examine this aspect in greater
detail in \S 5). A companion video of the simulation
corresponding to Figures \ref{fig4}(a)-(c) is shown in Video1.mpg.
In addition to the temperature perturbation $\Delta T$, we also
display velocity vectors (shown by arrows). The animation enables one
to discern the separation of the modes in the strongly magnetized
region ($\beta \ll 1$). The transverse MHD wave can be easily
identified in the central sections of the tubes. The first front of
this wave crosses the $\beta=1$ contour at about 70 s and is speeded
up as it moves into the higher atmosphere, reaching the top boundary
at about 105 s. On the other hand, the slow magnetacoustic wave (for
which the velocity is principally aligned with the magnetic field)
reaches the top boundary at about 180 s. The final frame of the video
at $t=174$~s clearly shows the presence of localized shock-like
features with strong temperature enhancements of about 3000 K and
flows close to the sound speed. However, the temperature perturbations
and flows in the ambient (weak field) medium are very small.

It is instructive to compare the velocity pattern for the
configurations treated in the previous two cases in order to identify
the nature of the waves in different regions of the atmosphere. We
decompose the flow into components along the field ($V_s$) and normal
to it ($V_n$).  Positive (negative) values of $V_n$ correspond to the
velocity component parallel (anti-parallel) to $x$-axis when the
field is vertical. Figures \ref{fig5}a and \ref{fig5}b show the field
aligned ($V_s$) and transverse ($V_n$) components of the velocity for
a single tube at $t = 120$ s for the same parameters as before. We
only consider regions with a field strength greater than 20~G (for
weak fields the decomposition into longitudinal and transverse
components is not meaningful). 

Figures \ref{fig5}c and \ref{fig5}d show similar diagrams
for the model with multiple tubes. In Figs.~\ref{fig5}a and
\ref{fig5}c we see an antisymmetric (about the axis) velocity pattern
in the strong field region ($\beta \ll 1$); this pattern can be
identified with the longitudinal acoustic mode. On the other hand, in
Figs.~\ref{fig5}b and \ref{fig5}d we see a symmetric pattern as a
uniform halo, corresponding to the transverse MHD mode.  There is a
clear separation of the modes above the $\beta = 1$ level, where the
fast mode speed increases rapidly with height. The associated increase
in wavelength manifests itself as an increase in the spacing of the
color contours. The shape of the contours is a consequence of the
Alfv\'{e}n speed being maximum on the axis (at a given height).
Figures \ref{fig6}a and \ref{fig6}b give a close up of the temperature
perturbation $\Delta T$ and velocity pattern near the central region
of the tube, corresponding to the multiple and single tube
configurations, respectively. Both cases clearly show the change in
the flow pattern close to the $\beta = 1$ level and the anti-symmetric
structure of the acoustic waves away from the tube axis.

We find that the wave patterns in the single and multiple tube cases
are qualitatively similar. In both cases the acoustic and
magnetic type of modes are present.  Even though the initial
perturbation at the base ($z = 0$) is transverse, strong field aligned
motions develop particularly close to the central region of the tube,
similar to the finding in Paper I for a single tube, which suggests
that this is a general feature of the network.

\section{Results for Wave Generation in the External Medium}

So far we have examined the excitation of waves in flux tubes due to
transverse motions of their footpoints. We now investigate the
situation when the source of waves is in the field free medium, and
we consider how the waves interact with the flux tube in order to
assess how effectively they penetrate into flux tubes and heat the
network. Let us consider a localised source that generates acoustic
waves due to vertical periodic motions. The acoustic source is located
at the lower boundary of the model ($z = 0$) and has a Gaussian
profile in $x$ with a $1/e$-width of 50 km. As before, we take the
amplitude of the motions as 750 m s$^{-1}$ and period as 24~s. As we
shall see the nature of wave excitation in the tube depends on the
``angle of attack'' or the angle between the direction of the wave
front and a field line. We treat two cases corresponding to distant
and near acoustic sources on the lower boundary.
 
First consider an acoustic source at $x = 2.3$~Mm, which is 500 km
away from the axis of the flux rope. Figures \ref{fig7}(a)-(c) show
the temperature perturbations at different epochs, corresponding to
$t=76$~s, 101~s and 140~s respectively. It is seen
that the acoustic wave strikes the tube ($\beta \approx 1$ surface)
almost normal to the field and generates a fast MHD wave in the tube;
for a more thorough description of what happens at this interface,
see the lower panel in Fig.~3 of \citet{Cally2007}. There is also slow
magnetoacoustic wave generated that essentially travels along the
field lines.  Figures \ref{fig7}(b)-(c) shows the distortion of the
wave front as the wave propagates into the low $\beta$ region and
where it is refracted upwards (away from the tube axis).  A more
informative way to examine the nature of the waves is to plot the
longitudinal and transverse velocity components, which is shown for
the present case in Figures \ref{fig8}(a)-(b) at time $t = 140$~s
(for $B > 20$~G). The field aligned component $V_s$, that essentially
depicts the acoustic wave, does not penetrate significantly into the
central regions of the tube as we see from Figure \ref{fig8}(a). On
the other hand, the normal component $V_n$, associated with the fast
MHD mode (where the $\beta < 1$) dominates in the central regions of
the tube. This wave is generated due to mode conversion when the
acoustic wave strikes the $\beta = 1$ surface.

Now consider an acoustic source at $x = 1.9$~Mm, only 100 km from the
flux tube axis. The vertical field $B_z (x,0)$ at the base has a
nearly Gaussian profile with $1/e$-width of 65 km, and the acoustic
source has $1/e$-width of 50 km, so there is significant overlap
between the two profiles. Figures \ref{fig7}(d)-(f) show the
temperature perturbation for $t=76$~s, 101~s and 140~s, respectively.
In contrast to the previous case, such a source excites waves that
travel with almost no distortion by the flux tube, as can be seen
by the fact that the wave front remains close to circular. From
Figures \ref{fig7}(e) we find that the angle between the wave vector
and the field lines is very small at the $\beta = 1$ level. In this
case, there is only a small amount of mode conversion into a
transverse magnetic mode. In the magnetized region ($B > 20$~G), the
waves are dominantly field-aligned, as can be seen in Figures
\ref{fig8}(c)-(d) at time $t = 140$~s, which show the field-aligned
and normal components of the velocity. In the region with $\beta < 1$
these waves are slow magnetoacoustic modes, and the fast mode has much
less power.

\section{Summary and Discussion}

\begin{figure} 
\epsscale{0.9}
\plotone{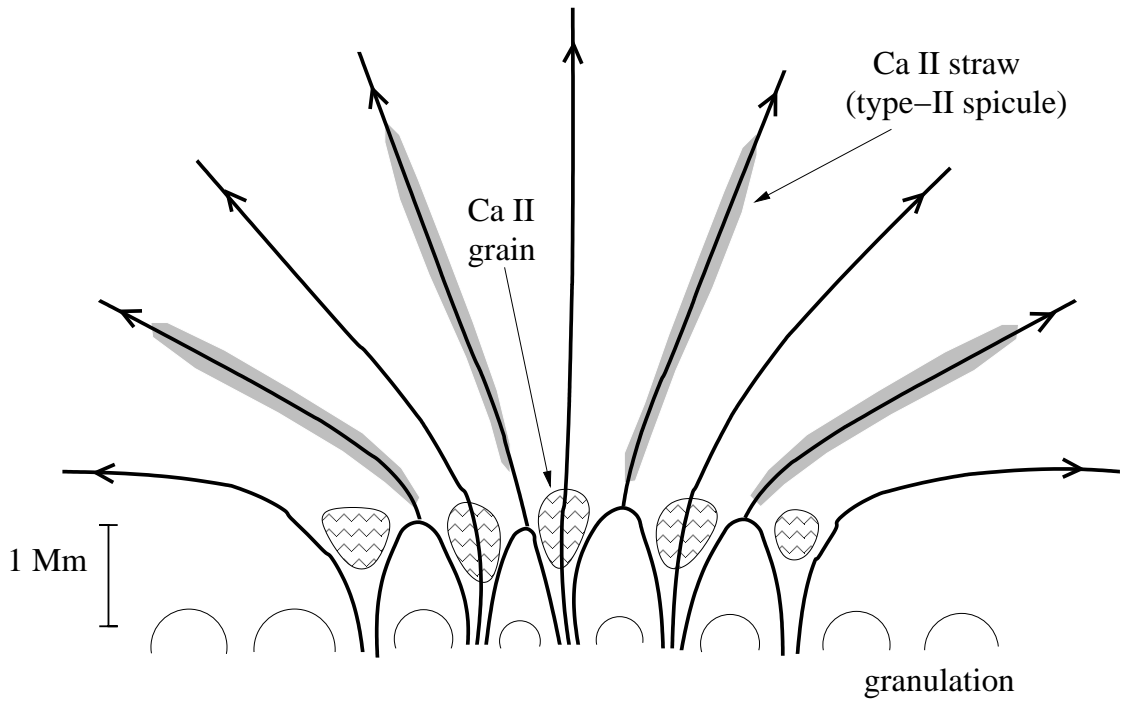}  
\caption
{Schematic diagram showing the structure of a magnetic network element
on the quiet Sun. The thin half-circles at the bottom of the figure
represent the granulation flow field, and the thick curves represent
magnetic field lines of flux tubes that are rooted in the
intergranular lanes. The Ca II bright grains are thought to be located
inside the flux tubes at heights of about 1 Mm above the base of the
photosphere. We suggest that the Ca II ``straws'' \citep[][]
{Rutten2006} may be located at the boundaries between the flux tubes.}
\label{fig2}
\end{figure}

\begin{figure} 
\epsscale{0.9}
%\hspace*{-0.7in}
\plotone{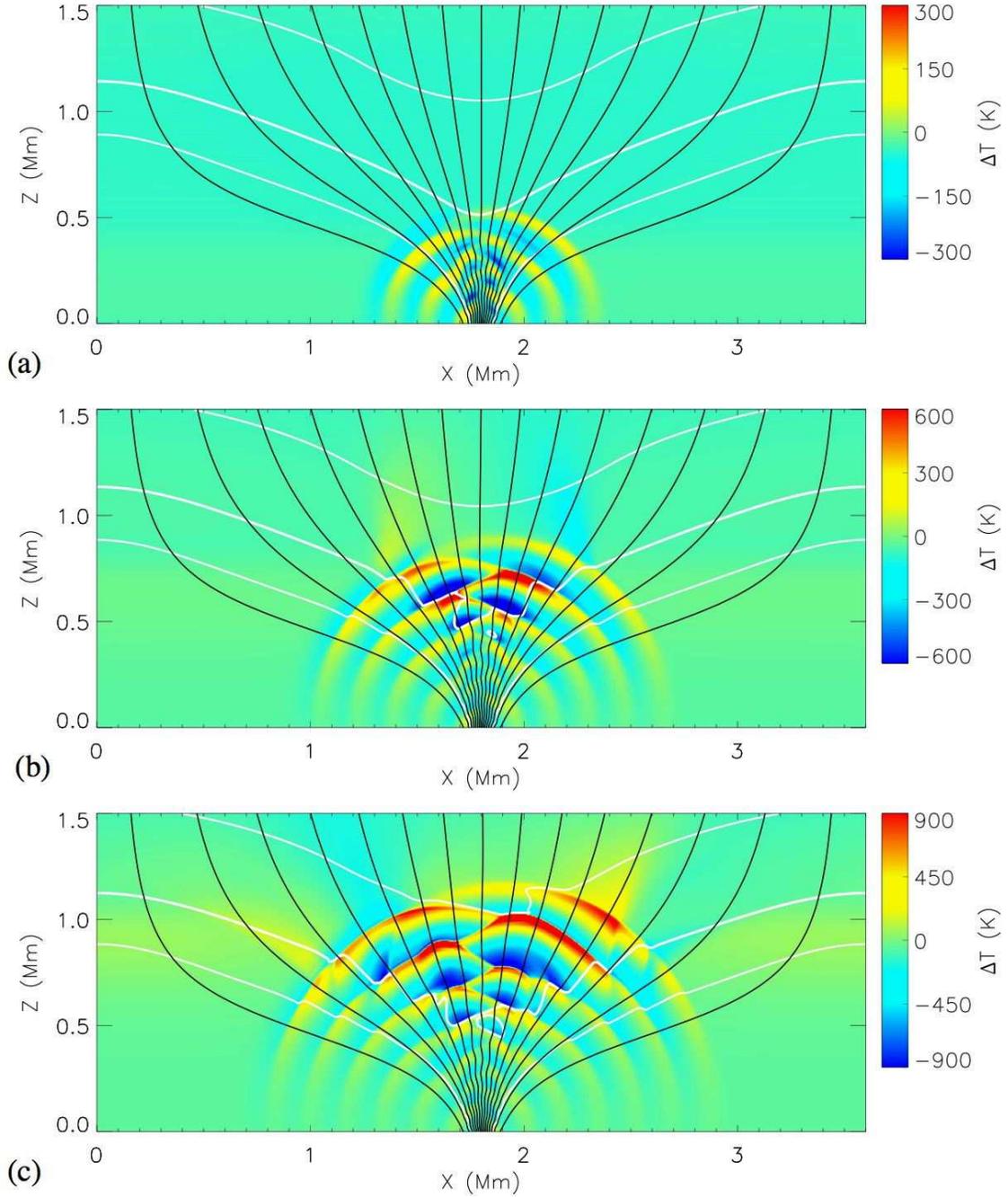}  
\caption
{The temperature perturbation, $\Delta T$, (about the initial state)
at (a) 75 s, (b) 122 s, and (c) 153 s in a network element due to
periodic horizontal motion at the lower boundary, with an amplitude
of 750 m s$^{-1}$, and a period of 24 s. The black curves denote
the magnetic field lines, and the color scale shows the temperature
perturbation.  The white curves denote contours of constant $\beta$
corresponding to $\beta=0.1$ (upper curve), 1.0 (thick curve) and 10
(lower curve).}
\label{fig3}
\end{figure}

\begin{figure} 
\epsscale{0.8}\hspace*{-0.7in}
\plotone{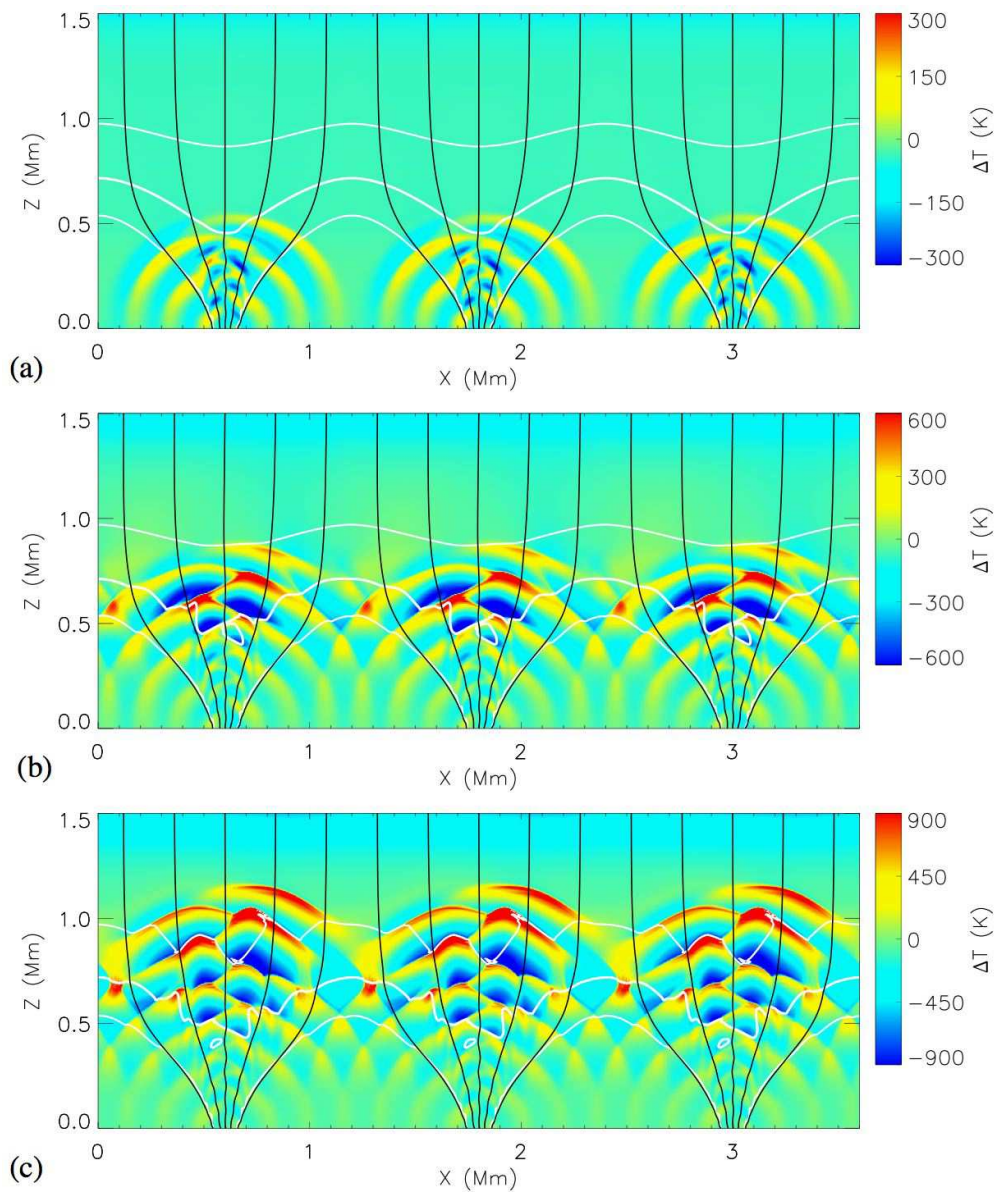}  
\caption
{The temperature perturbation, $\Delta T$, (about the initial state)
at (a) 75~s, (b) 122~s, and (c) 153~s in a network region consisting
of 3 flux tubes.  Wave excitation is due to periodic horizontal motion
at the lower boundary, with an amplitude of 750 m s$^{-1}$, and a
period of 24~s.  The black curves denote the magnetic field lines, and
the color scale shows the temperature perturbation.  The white curves
denote contours of constant $\beta$ corresponding to $\beta=0.1$
(upper curve), 1.0
(thick curve) and 10 (lower curve). This figure is also available as an mpeg animation
in the online version of the journal.}
\label{fig4}
\end{figure}

\begin{figure} 
\epsscale{0.65}\hspace*{-0.7in}
\plotone{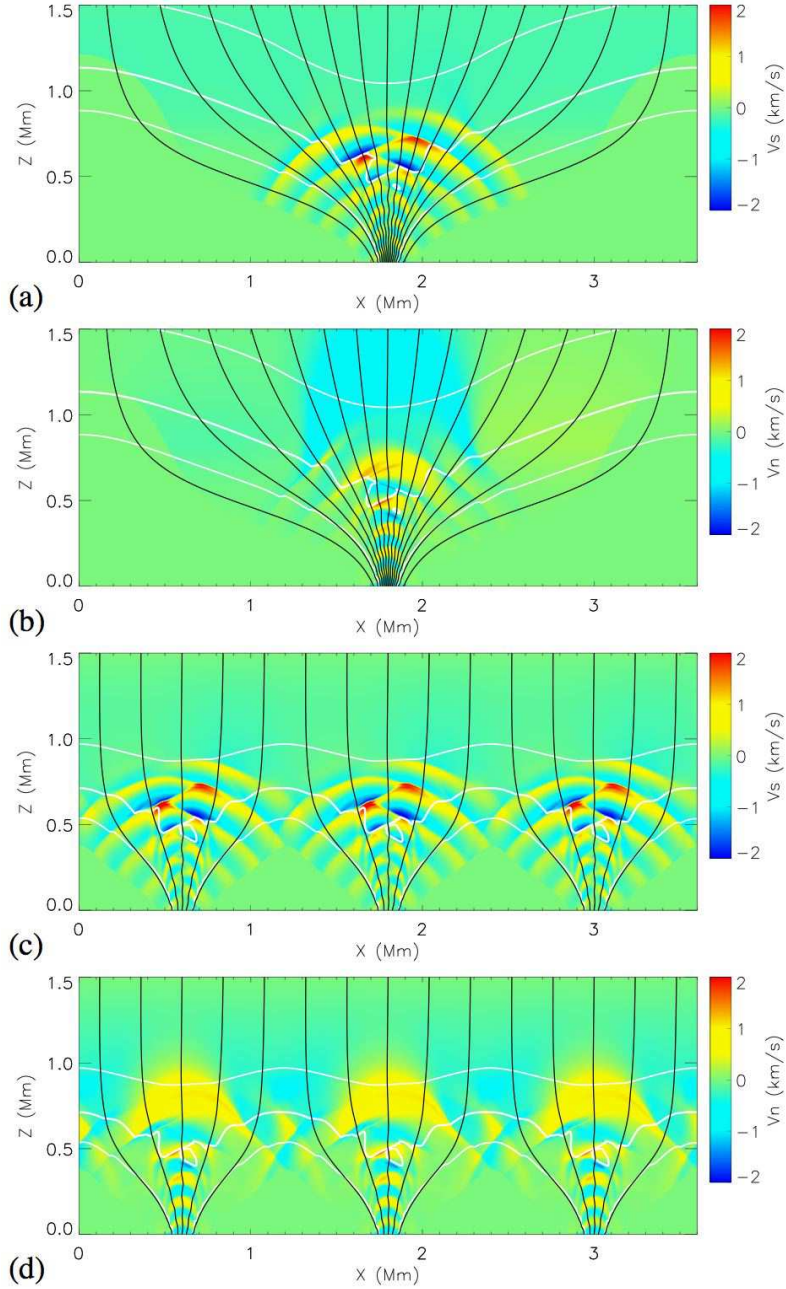}  
\caption
{Field aligned ($V_s$) and transverse ($V_n$) components of the
velocity in a single tube,  (panels {\bf a} - {\bf b}) and multiple
tubes (panels {\bf c} - {\bf d}) at $t = 120$~s due to periodic
horizontal motion at the lower boundary, with an amplitude of 750
m s$^{-1}$, and a period of 24~s. The black curves are magnetic field
lines, and the white curves are contours of constant $\beta$,
corresponding to $\beta=0.1$, 1.0 and 10. In the weak-field region
($B < 20$~G) the decomposition of velocity into field-aligned and
transverse components is ambiguous, and is not shown.}
\label{fig5}
\end{figure}

\begin{figure} 
\epsscale{1.0}\hspace*{-0.7in}
\plotone{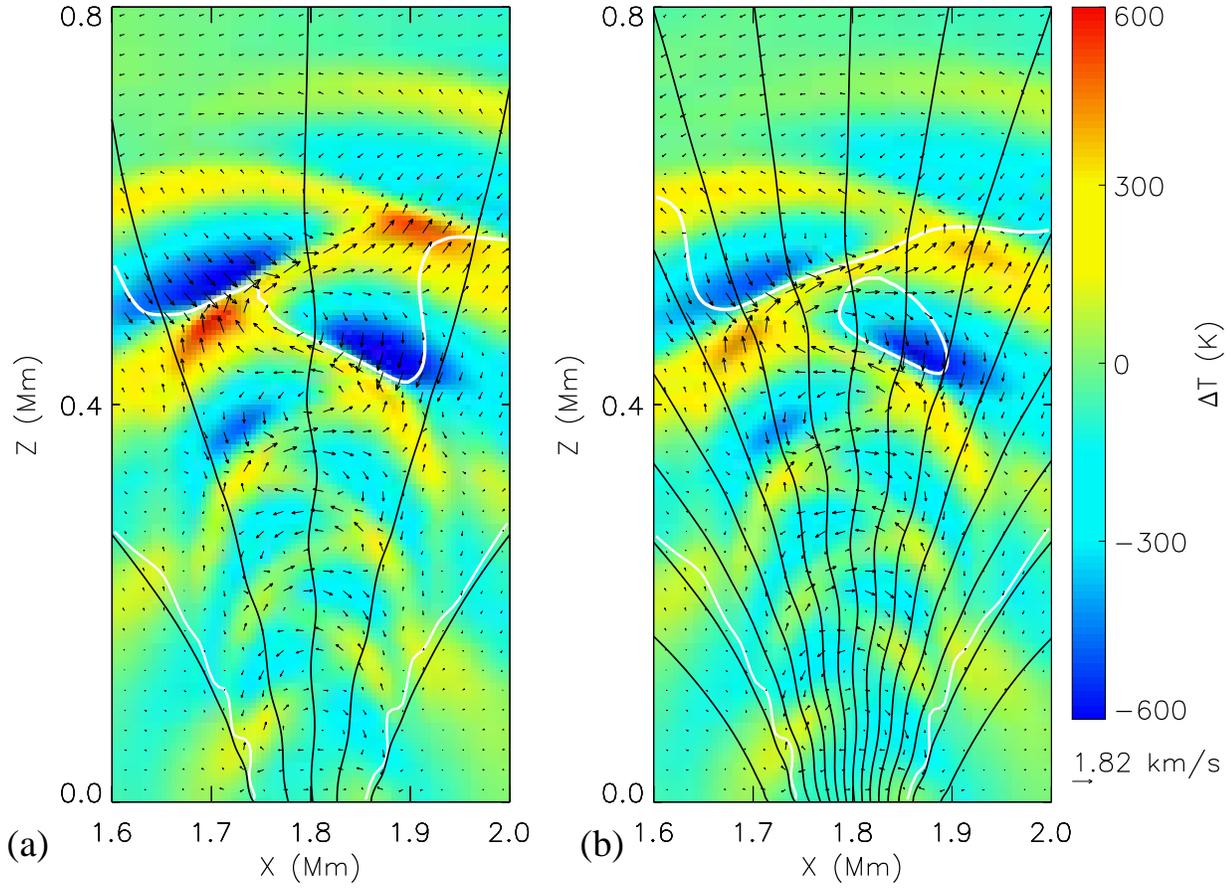}  
\caption
{Close-up of the temperature perturbation $\Delta T$ and velocity
field at time $t = 101$ s for (a) the central flux tube shown in
Fig.~\ref{fig4}, and (b) the single flux tube shown in
Fig.~\ref{fig3}. The black curves are field lines, and the white
curves are contours of plasma $\beta$ ({\it thick}: $\beta=1$;
{\it thin}: $\beta=10$).}
\label{fig6}
\end{figure}

\begin{figure} 
\epsscale{1.0}\hspace*{-0.7in}
\plotone{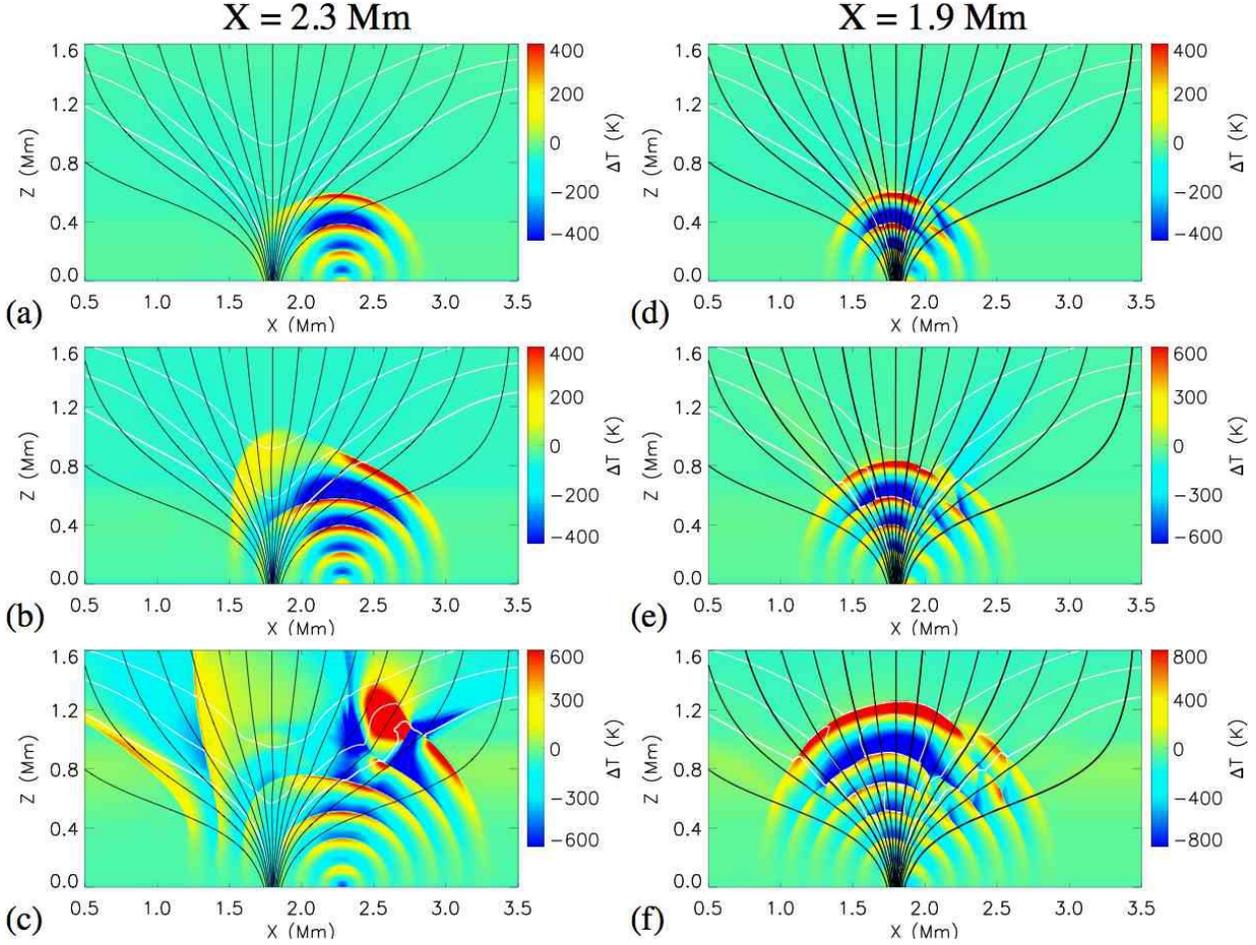}  
\caption
{Temperature perturbation, $\Delta T$, (about the initial state) at
(a) 76~s, (b) 101~s and (d) 140~s in a single tube due to a localised
vertical periodic motion at $z = 0$, $x = 2300$~km (panels
{\bf a}- {\bf c}) and $x = 1900$~km (panels {\bf d}- {\bf f}) with
an amplitude of 750 m s$^{-1}$, and a period of 24~s.  The black
curves denote the magnetic field lines, and the white curves are
contours of constant $\beta$, corresponding to $\beta=0.1$, 0.3 and
1.0.}
\label{fig7}
\end{figure}

\begin{figure} 
\epsscale{1.0}\hspace*{-0.7in}
\plotone{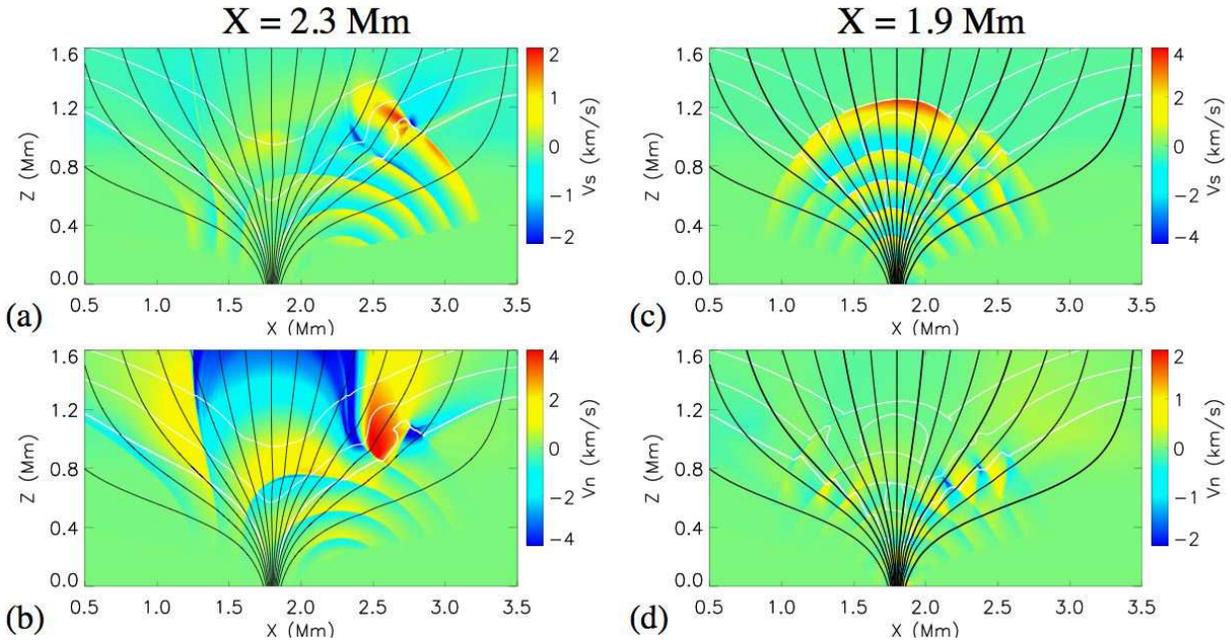}  
\caption
{Field-aligned ($V_s$) and transverse ($V_n$) components of the
velocity at $t = 140$ s for a localized acoustic source at $z = 0$,
$x = 2.3$~Mm (panels [a] and [b]) and $x = 1.9$~Mm  (panels [c] and
[d]). The black curves denote the magnetic field lines, and the white
curves are contours of constant $\beta$ ($\beta=0.1$, 0.3 and 1.0).
The velocity in the region corresponding to $B < 20$~G is not shown.}
\label{fig8}
\end{figure}

Recent observations of the chromospheric network were discussed. We
argued that Ca~II network grains represent plasma located at heights
between about 0.5 Mm and 1.5 Mm, and are distinct from the
spicule-like features observed at larger heights (see
Fig.~\ref{fig2}). The heating in network grains must occur in a
sustained (as opposed to intermittent) manner. The physical processes
involved in this heating are not well understood. Long-period waves
are thought to be involved in driving spicules \citep[e.g.,][]
{DePontieu2004}, but we argued against long-period waves as the cause
of the heating in Ca~II network grains. Instead, our hypothesis in
this paper is that the heating in network grains is caused by weaker
shocks that occur repetitively at short time intervals ($P < 100$ s)
to maintain the enhanced temperature of the network grains. The
weakness of the shocks is needed to explain the observed symmetrical
shapes of the Ca~II H \& K line profiles from network grains.
The symmetry of the line profiles and relative constancy of the Ca~II
emission may be explained if the shocks have small horizontal scales
(a few hundred km) and multiple shocks are superposed along the line
of sight. At present there is no observational evidence for such
short-period waves in network elements.

We studied the propagation of short-period magnetoacoustic waves
in magnetic elements, using the 2-D MHD model presented in Paper~I.
The waves are launched as transverse kink waves at the base of the
photospheric flux tubes. The present work complements Paper~I in two
important respects: (a) we consider an equilibrium model with lower
magnetic field strength, so that the $\beta = 1$ surface is located in
the upper photosphere; (b) we examine the interactions of waves from
neighboring tubes. We have identified an efficient mechanism for the
generation of longitudinal motions in the chromosphere from purely
transverse motions at the base of tubes. These longitudinal motions
are associated with slow magnetoacoustic waves that can contribute to
a localized heating of the chromosphere.

The qualitative features of the results for a single flux tube are
similar to those found in Paper I. Irrespective of whether the field
at the base of the flux tube is strong or weak, the transverse waves
at the base are converted at chromospheric heights into two sets of
slow-mode shocks that are 180$^\circ$ out of phase with each other and
propagate upward along the magnetic field lines on opposite sides of
the flux tube axis. We propose that network grains are heated by
dissipation of such shocks. This heating occurs mainly in the central
part of the flux tube. We also considered an equilibrium configuration
with multiple tubes, and found that through the interference of waves
emanating from neighboring tubes, there is some heating at the tube
interfaces. However, the dominant heating occurs near the tube axes.

For the case of a single flux tube we found that the width of the
shocked region at time $t = 153$ s is about 1200 km (see Fig.~3c),
a significant fraction of the assumed width the flux tube at large
height. For the case with multiple flux tubes, the width of the
shocked region is about 900 km, or 3/4 of the width of one flux tube
(see Fig.~4c). Therefore, the spatial extent of the
shocked region in our models is larger than the subarcsecond size of
the Ca~II H network grains observed by DOT (see Fig.~1c). Part of
the difference may be explained by the fact that the Ca~II H images
contain a significant photospheric contribution \citep[][]
{Rutten2006, Rutten2007}, which originates at lower heights where
the flux tubes are thinner. Another possible explanation lies in the
details of spectral line formation: as the shocks propagate to larger
height, the density behind the shocks decreases and the Ca~II H line
source function drops below the Planck function, producing the
``absoption'' feature at line center. Therefore, the chromospheric
contribution to the observed emission comes from a limited range of
heights, and may reflect only the lower part of the shocked region
shown in Figures 3c and 4c. To determine whether the proposed models
can explain the observed widths of network grains, more detailed
modeling of the formation of the Ca~II H line is needed.

We also investigated the interaction of a flux tube with acoustic
waves from the external medium. A careful investigation shows
that their penetration into the tube depends on the angle between the
wave front and field lines, or ``angle of attack" \citep[][and
references therein] {Carlsson2006, Cally2007}. \citet{Schunker2006}
and \citet{Cally2007} examined the interaction between an acoustic
wave and a magnetic field, and found that at the level where the sound
and Alfv\'en speeds are equal there is both mode {\em transmission}
and mode {\em conversion}. The former refers only to a nomenclature
change as a fast wave in the ambient medium is transmitted as a slow
wave in the flux tube.  However, the {\em transmitted} wave continues
to be acoustic in character.  On the other hand, in the latter case,
the wave entering the magnetic medium remains fast throughout and
changes character from acoustic in the ambient medium to magnetic in
the flux tube, which corresponds to a mode {\em conversion}.  We
examined this phenomenon by considering two limiting cases of distant
and near acoustic sources. Our calculations clearly show that acoustic
waves, generated in the ambient medium, do not significantly penetrate
inside the flux tube unless the acoustic source is located in its
immediate vicinity (i.e., within about one flux tube radius).
Consequently, short-period waves from the external medium do not
offer a promising mechanism for heating the magnetic network at
heights below 1500 km.

Future observations with Hinode/SOT should be able to establish
whether short-period magnetoacoustic waves exist in flux elements at
the heights where the Ca~II ${\rm H}_{2V}$ and ${\rm H}_{2R}$ emission
features are formed, and if so, whether they play a significant role
in chromospheric heating. If such waves are not found, Alfv\'{e}n
waves may provide an alternative, but such waves may be difficult to
detect in solar disk observations. Future modeling should consider the
interactions of waves with magnetic flux tubes in three dimensions in
order to allow for the coupling between all three MHD wave modes.
The models should include a more realistic energy equation than the
one considered here. Both p-mode waves and waves generated locally
at the base of the flux tubes should be considered. Based on such
models, Ca~II H and K profiles should be simulated to determine
whether the models can reproduce the observed symmetrical profiles
from network grains.

\acknowledgments
We thank the referee for detailed comments on an earlier draft of this
paper. The Dutch Open Telescope is operated by Utrecht University at
the Spanish Observatorio del Roque de los Muchachos of the Instituto
de Astrofisica de Canarias.

\clearpage

\begin{deluxetable}{llllll}
\tablecolumns{6}
\tablewidth{0pc}
\tablecaption{Parameters on the Tube Axis \& Ambient Medium }
\tablehead{
\colhead{}    &  \multicolumn{2}{c}{Tube Axis} &   \colhead{}   &
\multicolumn{2}{c}{Ambient Medium} \\
\cline{2-3} \cline{5-6} \\
\multicolumn{1}{l}{Variable} & \multicolumn{1}{l}{$z = 0$ } & \multicolumn{1}{l}{$z = 1500$ km} &
        \colhead{} & \multicolumn{1}{l}{$z = 0 $} & \multicolumn{1}{l}{$z = 1500$ km}  } 
\startdata
Temperature & 4700 K & 8700 K & & 4700 K & 8700 K  \\
Density & 2.6 10$^{-7}$ g cm$^{-3}$ & 2.1 10$^{-12}$ g cm$^{-3}$ & & 
          3.9 10$^{-7}$ g cm$^{-3}$ & 3.2 10$^{-12}$ g cm$^{-3}$  \\
Pressure & 8.0 10$^{4}$ dyn cm$^{-2}$ & 2.2  dyn cm$^{-2}$ & & 
          1.2 10$^{5}$ dyn cm$^{-2}$ & 3.3  dyn cm$^{-2}$  \\
Sound speed & 7.1 km s$^{-1}$  & 13 km s$^{-1}$  &  
            & 7.1 km s$^{-1}$  & 13 km s$^{-1}$  \\ 
Alfv\'en speed & 5.6 km s$^{-1}$  & 85 km s$^{-1}$  & & \nodata & \nodata \\ 
Magnetic field & 1000 G  & 44 G & & \nodata & \nodata  \\
$\beta$ & 2.0  & 2.8 10$^{-2}$ & & \nodata & \nodata  \\
\enddata
\end{deluxetable}

\clearpage

\clearpage

\clearpage

\clearpage

\clearpage

\clearpage

\clearpage

\clearpage

\end{document}